\documentclass[twocolumn,showpacs,prl,amsmath,amssymb]{revtex4}
\usepackage{graphicx}
\usepackage{bm}
\usepackage{color}

\begin{document}

\title{Theory of spin-orbit enhanced electric-field control of magnetism in multiferroic BiFeO$_3$} 
\author{Rog\'{e}rio de Sousa}
\email[]{rdesousa@uvic.ca} \author{Marc Allen} \affiliation{Department
  of Physics and Astronomy, University of Victoria, Victoria, B.C.,
  V8W 3P6, Canada} \author{Maximilien Cazayous}
\affiliation{Laboratoire Mat\'eriaux et Ph\'enom\`enes Quantiques (UMR
  7162 CNRS), Universit\'e Paris Diderot-Paris 7, 75205 Paris cedex
  13, France}

\date{\today}

\begin{abstract}
  We present a microscopic theory that shows the importance of
  spin-orbit coupling in perovskite compounds with heavy ions.  In
  BiFeO$_3$ (BFO) the spin-orbit coupling at the bismuth ion sites
  results in a special kind of magnetic anisotropy that is linear in
  the applied $E$-field. This interaction can convert the cycloid
  ground state into a homogeneous antiferromagnet, with a weak
  ferromagnetic moment whose orientation can be controlled by the
  $E$-field direction.  Remarkably, the $E$-field control of magnetism
  occurs without poling the ferroelectric moment, providing a pathway
  for reduced energy dissipation in spin-based devices made of
  insulators.
\end{abstract}



\maketitle


The ability to control magnetism using electric fields is of great
fundamental and practical interest. It may allow the development of
ideal magnetic memories with electric write and magnetic read
capabilities \cite{scott07}. The traditional mechanism of $E$-field
control of magnetism is based on the dependence of magnetic anisotropy
on the filling of $d$-orbitals. This allows $E$-field control of
magnetism in metallic materials such as magnetic semiconductors
\cite{chiba08} and ferromagnetic thin films \cite{shiota12}, but not
in insulators.  A method to influence magnetism using $E$-fields in
insulators is desirable because it would not generate electric
currents, potentially allowing the design of spin-based devices with
much lower energy dissipation \cite{rovillain10}.

\begin{figure}
\includegraphics[width=0.5\textwidth]{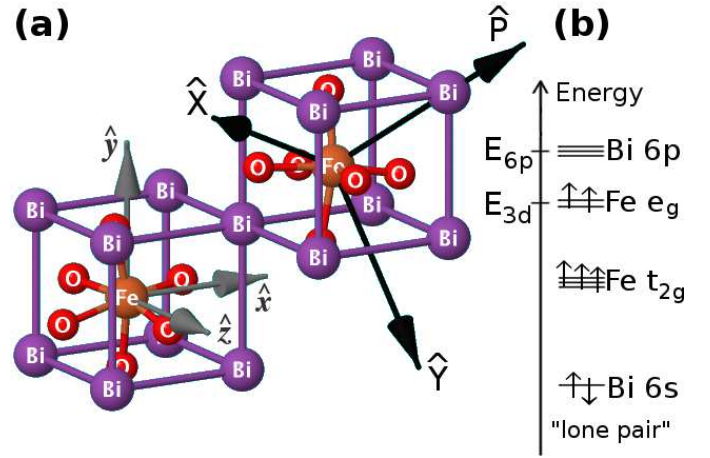}
\caption{(color online) (a) Conventional unit cell for BFO.  The
  simple cubic axes $\bm{\hat{x}}$, $\bm{\hat{y}}$, $\bm{\hat{z}}$ are
  denoted by grey vectors.  The ferroelectric polarization $\bm{P}$ is
  shown pointing along [111], arising mostly from the displacement of
  Bi ions with respect to the oxygens.  The directions $\bm{\hat{X}}$
  and $\bm{\hat{Y}}$ denoted by black vectors describe the plane
  perpendicular to $\bm{P}$ \cite{noteaxes}.  (b) Energy level diagram
  for Fe$^{3+}=[{\rm Ar}]3d^5$ and Bi$^{3+}=[{\rm Pt}]6s^2$ orbitals in
  BFO; the Fermi level lies between $E_{3d}$ and $E_{6p}$.}
\label{fig1}
\end{figure}

In insulators, the interactions that couple spin to electric degrees
of freedom, the so called magnetoelectric interactions, are usually
too weak to induce qualitative changes to magnetic states. A
remarkable exception occurs in the presence of the linear
magnetoelectric effect (LME), an interaction that couples spin and
charge linearly in either the external electric field $\bm{E}$ or the
internal electric polarization $\bm{P}$ of the material. Multiferroic
insulators with coexisting magnetic and ferroelectric phases have
emerged as the natural physical system to search for LME and enhanced
cross correlation between electricity and magnetism \cite{spaldin10}.
In a large class of multiferroic materials, the dominant form of LME
was found to be due to the spin-current effect (a type of
Dzyaloshinskii-Moriya interaction \cite{katsura05}), that couples
localized spins according to ${\cal H}_{{\rm SC}}=\sum_{i<j}J_{{\rm
    SC}}(\bm{P}\times \bm{R}_{ij})\cdot (\bm{S}_{i}\times\bm{S}_{j})$,
with $\bm{R}_{ij}$ the vector linking the atomic location of spin
$\bm{S}_i$ to the atomic location of spin $\bm{S}_j$.  In
manganese-based multiferroics, the spin-current interaction leads to
magnetic induced ferroelectricity and thus allows magnetic field
control of ferroelectricity \cite{kimura03}.

For $E$-field control of magnetism, research has been centered instead
on iron-based multiferroics, with bismuth ferrite [BiFeO$_3$ or BFO]
being the most notable example \cite{catalan09}. At temperatures below
$1143$~K, BFO develops a strong electric polarization
$P=100$~$\mu$C/cm$^2$ that points along one of the eight cube
diagonals of its unit cell [Fig.~\ref{fig1}(a)]. It becomes an
antiferromagnet below $643$~K with Fe spins forming a spiral of the
cycloid type, described by antiferromagnetic N\'{e}el vector
$\bm{\hat{L}}=\sin{(\bm{q}\cdot \bm{r})}\bm{\hat{q}}+\cos{(\bm{q}\cdot
  \bm{r})}\bm{\hat{P}}$.  The microscopic origin of the cycloid can
also be understood as arising from the spin-current interaction
\cite{rahmedov12}.  Plugging the cycloid $\bm{L}$ into ${\cal H}_{{\rm
    SC}}$, one finds that the lowest energy state is always achieved
when the cycloid wavevector $\bm{q}$ is perpendicular to $\bm{P}$.
Hence the spins are pinned to the plane formed by $\bm{P}$ and
$\bm{q}$. This fact has been central to all demonstrations of
$E$-field control of magnetism in multiferroics published to date; the
application of an $E$-field poles $\bm{P}$ from one cube diagonal to
another, forcing the spin cycloidal arrangement to move into a
different plane \cite{lebeugle08,heron11}.

In addition, BFO is known to have a weak magnetization $\bm{M}\propto
\bm{\hat{Z}}\times \bm{L}$ that is generated by an additional
Dzyaloshinskii-Moriya interaction \cite{ederer05,desousa09} (with
$\hat{\bm{Z}}\parallel \hat{\bm{P}}$).  When $\bm{L}$ is in a
cycloidal state, $\bm{M}$ is sinusoidal and averages out to zero over
distances larger than the cycloid wavelength.  The amplitude of the
oscillatory $|M|=0.06\mu_B/{\rm Fe}$ was measured recently
\cite{ramazanoglu11b}.  Therefore, a method to convert the cycloid
into a homogeneous state would preclude $\bm{M}$ from averaging out to
zero, with potential applications to electrically-written magnetic
memories.

A recent experiment \cite{rovillain10} suggested that the
spin-current interaction is not the only LME present in BFO.  The
application of an external $E$-field to bulk BFO was shown to result
in a giant shift of magnon frequencies \emph{that was linear in $\bm{E}$} and
$10^{5}$ times larger than any other known $E$-field effect on magnon
spectra.  

In this letter, we present a microscopic theory of $E$-field induced
magnetic anisotropy, and argue that it can provide an effective source
of LME in insulators with large spin-orbit coupling.
Our predicted LME explains the origin of the $E$-field effect on
magnon spectra measured in bulk BFO \cite{rovillain10}. Moreover, we
show that this effect is capable of switching the cycloidal spin state
of BFO into a homogeneous magnetic state with its orientation tunable
by the direction of the applied $E$-field.

{\it Model and microscopic calculation of LME.---}Our microscopic
Hamiltonian for coupling between spin and electric-field has three
contributions, ${\cal H}={\cal H}_{{\rm latt}}+{\cal H}_{{\rm
    elec}}+{\cal H}_{{\rm SO}}$. The first term is due to the lattice,
${\cal H}_{{\rm latt}}=-\hbar^{2}\nabla^{2}/(2m_e) + V_{{\rm
    crystal}}(\bm{r})$, with $m_e$ and $\bm{r}$ the free electron mass
and its coordinate, respectively.  The second term is called
``electronic'', ${\cal H}_{{\rm elec}}=-e\bm{r}\cdot \bm{E}$, with
$e<0$ the electron's charge. The last and most important term is the
spin-orbit interaction, ${\cal H}_{{\rm SO}}=\zeta \bm{\ell}\cdot
\bm{\sigma}$, with $\bm{\ell}=-i\bm{r}\times\bm{\nabla}$ the
electron's orbital angular momentum and $\bm{\sigma}/2$ its spin
operator. We take the spin-orbit interaction to be dominated by the
heaviest ion of the lattice, and take bismuth in BFO as a prototypical
example.

Single ion anisotropy is known to arise as a correction to the total
spin energy that is second order in the spin-orbit interaction
\cite{abragam70}. In our case the largest contribution arises in the
fourth order of our total Hamiltonian, i.e., second order in ${\cal
  H}_{{\rm SO}}$ and second order in ${\cal H}_{{\rm latt}}$ or ${\cal
  H}_{{\rm elec}}$. An explicit calculation yields
\begin{equation}
{\cal H}_{SIA}=-\frac{1}{(2S)^{2}}\bm{S}\cdot \left[ \sum_{m,n} \frac{\bm{V}_{mn}\otimes \bm{V}_{nm}}{E_{6p}-E_{3d_{m}}}\right]\cdot \bm{S},
\label{hsia}
\end{equation}
where the spin operator $\bm{S}=\sum_{i=1}^{5}\bm{\sigma}_i/2$
represents all five electron spins in the Fe$^{3+}$ $d$-shell.
The numerator of Eq.~(\ref{hsia}) is an outer product between vectors
\begin{equation}
\bm{V}_{mn}=-\sum_{n',\bm{R}_{{\rm Bi}}}\frac{\langle 3d_{m}\mid {\cal H_{{\rm latt,~elec}}}\mid6p_{n'}\rangle
\langle 6p_{n'}\mid \zeta\bm{\ell}\mid 6p_{n}\rangle}{E_{6p}-E_{3d_{m}}},
\label{vnm}
\end{equation}
involving 6p and 3d localized orbitals at Bi and Fe, respectively,
with a sum over all vectors $\bm{R}_{{\rm Bi}}$ linking the central Fe
to each of its eight neighboring Bi. We evaluate Eq.~(\ref{vnm}) by
taking as Bi orbitals the states $|6p_x\rangle$, $|6p_y\rangle$,
$|6p_z\rangle$, and as Fe orbitals the $e_g$ states
$|3d_{3z^{2}-r^2}\rangle$ and $|3d_{x^{2}-y^{2}}\rangle$, written with
respect to the cubic axes $\bm{\hat{x}},\bm{\hat{y}},\bm{\hat{z}}$ of
BFO's parent perovskite lattice. The Fe $t_{2g}$ states are not
considered here because they are about $2$~eV lower in energy
\cite{abragam70}, and thus only give a small correction to
Eq.~(\ref{hsia}).

We now turn to an explicit evaluation of the matrix elements appearing
in Eq.~(\ref{vnm}).  The spin-orbit matrix element is given by
\begin{equation}
\langle6p_{n'}|\zeta\bm{\ell}|6p_{n}\rangle 
= - i\eta \;\bm{\hat{n}'}\times\bm{\hat{n}},
\end{equation}
with $\eta=0.86$~eV chosen to match the spin-orbit splitting measured
in isolated Bi ions \cite{noteeta}.  Using symmetry, all lattice
matrix elements $\langle 3d_{m}\mid {\cal
  H_{{\rm latt}}}\mid6p_{n'}\rangle$ can be expressed in terms of the
direction cosines of $\bm{R}_{{\rm Bi}}$ plus only two parameters:
$V_{pd\sigma}=\langle 3d_{3z'^{2}-r^{2}}|{\cal
  H}_{{\rm latt}}|6p_{z'}\rangle$ and $V_{pd\pi}=\langle
3d_{x'z'}|{\cal H}_{{\rm latt}}|6p_{x'}\rangle$, with $z'$ pointing
along $\bm{R}_{{\rm Bi}}$. 
A similar procedure can be applied to the electronic matrix elements
$\langle 3d_{m}\mid {\cal H_{{\rm elec}}}\mid6p_{n'}\rangle$, reducing
them to expressions that depend on the direction cosines of
$\bm{R}_{{\rm Bi}}$ plus matrix elements like $\langle
3d_{x'z'}|z'|6p_{x'}\rangle$, etc. 

In order to compute the vectors in Eq.~(\ref{vnm}), we need to sum over all Bi neighbors forming a
distorted cube around each Fe. We do this by converting the sum into an angular
integral, 
\begin{eqnarray}
\sum_{\bm{R}_{{\rm Bi}}}\langle 3d_{m}\mid {\cal H_{{\rm latt,~elec}}}&\mid&6p_{n'}\rangle \approx
\frac{8}{4\pi}\int d\Omega_{R}\left[
1+\delta\bm{R}\cdot \bm{\nabla_{R}}\right]\nonumber\\
&&\times\langle 3d_{m}\mid {\cal H_{{\rm latt,~elec}}}\mid 6p_{n'}\rangle,
\label{hybapprox}
\end{eqnarray}
with $\delta
\bm{R}=\left(R_{\parallel}\bm{\hat{P}}+u_{\perp}\bm{E}_{\perp}\right)$
denoting the deviation of the Bi ions from the perfect cube. This
includes Bi displacement along $\bm{\hat{P}}$ causing
ferroelectricity; the displacement is given by $R_{\parallel}=0.116
R_{{\rm Bi}}$ with $R_{{\rm Bi}}=4.88$~\AA~\cite{kubel90}.  The
component of $\bm{E}$ along $\bm{\hat{P}}$ can be neglected (it can
not compete with the internal field generated by ferroelectricity), so
we write the external $E$-field as
$\bm{E}_{\perp}=E_{\perp}[\cos{(\psi)\bm{\hat{X}}}+\sin{(\psi)}\bm{\hat{Y}}]$,
with the rhombohedral axes $\bm{\hat{X}}$ and $\bm{\hat{Y}}$ defined
in \cite{noteaxes} and shown in Fig.~\ref{fig1}(a). This perpendicular
component induces additional lattice displacement
$u_{\perp}E_{\perp}$; an estimate based on infrared spectroscopy
\cite{lobo07} yields $u_{\perp}E_{\perp}/R_{{\rm Bi}}=2.4 \times
10^{-4} E_{\perp}/(10^{5}{\rm V/cm})$.

After computing the averages over all matrix elements Eq.~(\ref{hsia}) yields
\begin{equation}
{\cal H}_{2}=-\frac{a}{2} \left(\bm{S}\cdot \bm{\hat{P}}\right)^{2}\label{aint},
\end{equation}
\begin{eqnarray}
{\cal H}_E&=&\frac{(\xi E_{\perp})}{2}
\left[\cos{\left(\psi\right)}S^{2}_{x}+\cos{\left(\psi-\frac{2\pi}{3}\right)}S^{2}_{y}\right.\nonumber\\
&&\left.+\cos{\left(\psi-\frac{4\pi}{3}\right)}S^{2}_{z}\right].\label{xiint}
\end{eqnarray}
Equation~(\ref{xiint}) depends linearly on $E_{\perp}$, i.e., it
gives rise to the LME.

Even in the absence of an external $E$-field, we find a
magnetic anisotropy,
\begin{equation}
a=\frac{1792\eta^{2}}{9(2S)^{2}}\frac{V^{2}_{\parallel}}{\left(E_{6p}-E_{3d}\right)^{3}},
\end{equation}
with a coupling energy related to the lack of inversion symmetry along
$\bm{P}$: $V_{\parallel}=(R_{\parallel}/R_{{\rm
    Bi}})(V_{pd\sigma}+V_{pd\pi}/\sqrt{3})$.

Taking $(E_{6p}-E_{3d})$ to be equal to BFO's band gap of $2.8$~eV
\cite{kumar08}, and using the tabulated values for the Fe-Bi bond
$V_{pd\sigma}=-71$~meV and $V_{pd\pi}=-41$~meV \cite{harrison04}, we
get $V_{\parallel}=-11$~meV and $a=32$~$\mu$eV$\doteq 0.4$~K.

The effect of the external $E$-field is to introduce magnetoelectric
coupling with reduced symmetry; from Eq.~(\ref{vnm}) we separate 
electronic and lattice contributions. 
The electronic LME is given by 
\begin{widetext}
\begin{eqnarray}
\xi_{{\rm elec}}&=&\frac{8}{35}\left(\frac{a}{V_{\parallel}}\right)
e\Bigl(\langle 3d_{3z'^{2}-r^2}|z'|6p_{z'}\rangle+\sqrt{3}\langle 3d_{x'z'}|z'|6p_{x'}\rangle + 3\sqrt{3}\langle 3d_{y'z'}|y'|6p_{z'}\rangle
+ \frac{5}{\sqrt{3}}\langle 3d_{x'y'}|y'|6p_{x'}\rangle\nonumber\\
&&- \langle 3d_{3z'^{2}-r^{2}}|x'|6p_{x'}\rangle
+ \frac{1}{\sqrt{3}}\langle 3d_{x'^{2}-y'^{2}}|x'|6p_{x'}\rangle\Bigl),
\label{xielec}
\end{eqnarray}
\end{widetext}
while the lattice LME is 
\begin{equation}
\xi_{{\rm latt}}=-\frac{4\sqrt{2}}{7}\left(\frac{a}{V_{\parallel}}\right)
\left(V_{pd\sigma}+\frac{V_{pd\pi}}{\sqrt{3}}\right)\left(\frac{u_{\perp}}{R_{{\rm Bi}}}\right).\label{xilatt}
\end{equation}
Note how these are physically distinct mechanisms: The lattice
mechanism is proportional to $u_\perp E_\perp$, i.e., it arises from
$E$-field induced lattice displacement contained in ${\cal H}_{{\rm
    latt}}$. Plugging the tabulated values for $V_{pd\sigma}$ and
$V_{pd\pi}$ we get $\xi_{{\rm latt}}=-5\times 10^{-2}\;\mu{\rm
  eV}/(10^{5}{\rm V/cm})$. The electronic mechanism is instead related to
$E$-field induced atomic orbital admixture, and its matrix elements are not
tabulated. Assuming $\langle 3d |x'_{i}| 6p\rangle\sim R_{{\rm Bi}}$
we get an order of magnitude estimate of $\xi_{{\rm elec}}\sim
+30\;\mu{\rm eV}/(10^{5}{\rm V/cm})$. 

{\it Comparison to experiments.---}The experiment in
Ref.~\cite{rovillain10} discovered a strong dependence of magnon
frequencies on the external $E$-field, and used group theory to fit
two kinds of $E$-field induced anisotropy:
$F_1=-(\xi/4)(\bm{E}_{\perp}\cdot \bm{S})(\bm{S}\cdot\bm{\hat{P}})$,
and $F_2=-(\xi/4) \bm{E}_{\perp}\cdot
[(S^{2}_{Y}-S^{2}_{X})\bm{\hat{X}}+(2S_{X}S_{Y})\bm{\hat{Y}}]$. It was
shown that only $F_2$ would give rise to the observed linear in
$E_{\perp}$ magnon shift, and a fit of $\xi_{{\rm exp}}=+55\; \mu{\rm
  eV}/(10^{5}{\rm V/cm})$ with $a=0$ was established at $T=300$~K.  To
compare this result to our theory, we write our Eq.~(\ref{xiint}) in
the rhombohedral basis and get that it is equal to $F_2+2\sqrt{2}F_1$.
Thus our Eq.~(\ref{xiint}) can be expressed as a function of the two
anisotropy terms of Ref.~\cite{rovillain10} and explains the origin of
the interaction leading to electrical control of magnons in BFO.

Our calculations are also supported by the good agreement between our
calculated zero-field anisotropy energy $a=32$~$\mu$eV and the value
of $a\sim 10$~$\mu$eV extracted from neutron diffraction experiments
\cite{ramazanoglu11}. 

We find that Eq.~(\ref{xiint}) will dominate over other known
magnetoelectric couplings in BFO for $E$-fields in the practical range
($E_{\perp}<10^{7}$~V/cm); this is shown in the supplemental material
section \cite{suppnote}.

 {\it Electric-field control of magnetism.---}To find out whether
 our effect can be used to control magnetism using an external
 $E$-field, we incorporate Eq.~(\ref{xiint}) into the usual continuum free energy model for BFO
 \cite{sparavigna94,sosnowska95,desousa08},
\begin{eqnarray}
{\cal F}&=&\int d^{3}x \left\{-\frac{m'}{2}L^{2}+\frac{c'}{2}\sum_{\gamma=x,y,z}\left|\nabla L_{\gamma}\right|^{2}\right.\nonumber\\
&&-\alpha' \bm{P}\cdot \left[\bm{L}\left(\nabla\cdot\bm{L}\right)+\bm{L}\times\left(\nabla\times \bm{L}\right)\right]\nonumber\\
&&+\frac{(\xi'  E_{\perp})}{2}\left[\cos{\left(\psi\right)}L^{2}_{x}
+\cos{\left(\psi-\frac{2\pi}{3}\right)}L^{2}_{y}\right.\nonumber\\
&&\left.\left.+\cos{\left(\psi-\frac{4\pi}{3}\right)}L^{2}_{z}\right]\right\}.
\label{hdens}
\end{eqnarray}
%
%
Here $\bm{L}$ is the N\'{e}el vector, and the first and second terms
inside the brackets of Eq.~(\ref{hdens}) arise from the exchange
interaction between spins; the third term arises from the continuum
limit of the spin-current coupling, leading to $\alpha'=J_{{\rm
    SC}}(\Omega_0/2)^{5/3}/(2S\mu_B)^{2}$, with
$\Omega_0=124.32$~\AA$^{3}$ the unit cell volume in BFO.  This term
explains the origin of the cycloid in BFO when $c'q^2\approx 1$
\cite{rahmedov12,sparavigna94,sosnowska95}.  The fourth term is the
continuum limit of Eq.~(\ref{xiint}), with
$2\xi'=(\Omega_0\xi)/(2S\mu_B)^2$.

\begin{figure}
\includegraphics[width=0.5\textwidth]{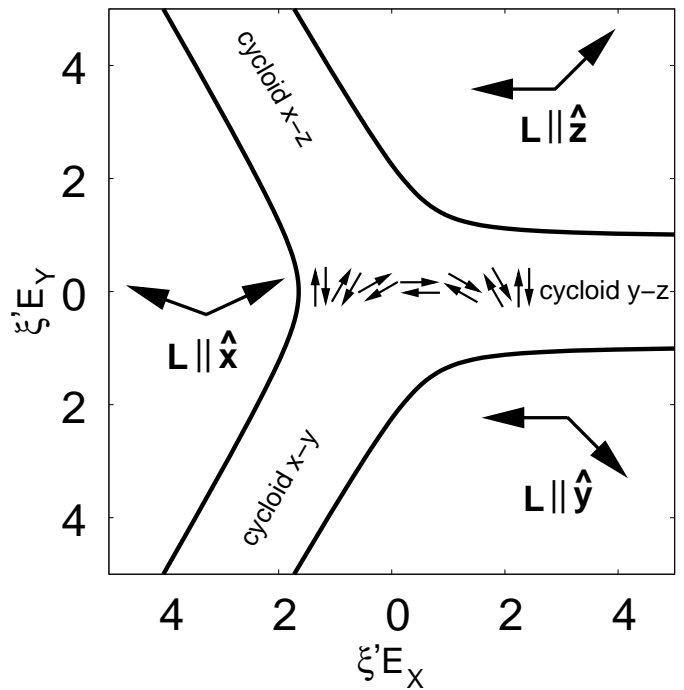}
\caption{Electric-field induced magnetic phase diagram for BFO. Here
  $E_X$ and $E_Y$ are projections of the external $E$-field in the
  plane perpendicular to $\bm{P}$, and $\xi'=1/(4\times 10^4 {\rm
    ~V/cm})$. The axes $\bm{\hat{X}}$, $\bm{\hat{Y}}$ are shown in
  Fig.~\ref{fig1}(a) \cite{noteaxes}. A transition from cycloid to
  homogeneous magnetism is predicted for certain directions in the
  $X-Y$ plane.  In this case the N\'{e}el vector $\bm{L}$ will point
  along one of the conventional cubic directions
  $\bm{\hat{n}}=\bm{\hat{x}},\bm{\hat{y}},\bm{\hat{z}}$, and due to
  the Dzyaloshinskii-Moriya coupling the magnetization $\bm{M}$ will
  point along $\bm{\hat{Z}}\times \bm{L}$. Thus an external $E$-field
  is able to control the direction of the two magnetic order
  parameters $\bm{M}$ and $\bm{L}$.}
\label{fig2}
 \end{figure}

The minimum free energy state $\bm{L}(\bm{r})$ can be found using
functional derivatives in the same way as done in
Refs.~\cite{sparavigna94,desousa08}.  The result is summarized in
Fig.~\ref{fig2}. At low $E_{\perp}$, the energy is minimized by a
cycloid with wavevector
$\bm{q}=(\alpha'P/c')[\sin{(\psi/2)}\bm{
  \hat{X}}+\cos{(\psi/2)}\bm{\hat{Y}}]$, lifting the cycloid
direction degeneracy of bulk BFO \cite{noteq}. As the electric
field is increased, the anisotropy energy favors an anharmonic cycloid
ground state with $\bm{L}$ forming a square wave along one of the
three cubic directions $\bm{\hat{x}}$, $\bm{\hat{y}}$, or
$\bm{\hat{z}}$, depending on the direction of $\bm{E}_{\perp}$. When
$E_{\perp}$ becomes larger than a certain critical value, we get a
phase transition to a homogeneous $\bm{L}$, effectively destroying the
cycloid state. The origin of this phase transition is the competition
between $E$-field induced anisotropy and the spin-current
interaction.  The free energy of the cycloid state is ${\cal F}_{{\rm
    cycloid}}\approx -(\xi'
E_{\perp}/2)\langle\cos^{2}{(\bm{q}\cdot\bm{r})}\rangle
-c'q^{2}/2=-\xi' E_{\perp}/4-c'q^2/2$.  Compare this to the free
energy of the homogeneous state, ${\cal F}_{{\rm Hom}}\approx
-\xi'E_{\perp}/2$; as $E_{\perp}$ increases, eventually we will have
$\xi'E_{\perp}>2c'q^2$ and ${\cal F}_{{\rm Hom}}<{\cal F}_{{\rm
    cycloid}}$, inducing a transition to the homogeneous state. Remarkably, the critical field is infinite when
$\bm{E}$ points antiparallel to one of the cubic directions. This is
also easy to understand from the symmetry of Eq.~(\ref{xiint}): for
example, when $\bm{E}_{\perp}$ points along the projection of
$-\bm{\hat{x}}$ in the $X-Y$ plane ($\psi=0^{\circ}$), the electric-field 
anisotropy energy is the same for $\bm{L}$ along $\bm{\hat{y}}$
or $\bm{\hat{z}}$; thus when $\bm{L}$ is a cycloid in the $y-z$ plane,
it is able to simultaneously minimize both the $E$-field anisotropy
and the spin-current energies; in this situation 
it is energetically favorable for $\bm{L}$ to remain a cycloid. 
A similar situation applies for $\bm{E}\parallel -\bm{\hat{y}}$
and $\bm{E}\parallel -\bm{\hat{z}}$.

 An important point is that $\bm{P}$ can be poled by the external
 $E$-field, changing the effective direction of $E_{\perp}$ in
 Fig.~\ref{fig2} (note that Fig.~\ref{fig2} assumes $\bm{P}\parallel
 [111]$ at all magnitudes of $\bm{E}_{\perp}$). To avoid poling, one
 can apply the $E$-field with the largest cube diagonal component
 along the [111] direction.  For example, using
 $\bm{E}=E[\cos{(30^{\circ})}\bm{\hat{Z}}-\sin{(30^{\circ})}\bm{\hat{X}}]$
 allows control of magnetism without changing $\bm{P}$, at the expense
 of having $E_{\perp}=E/2$.  Using $\xi_{{\rm
     exp}}=55$~$\mu$eV$/(10^{5}{\rm V/cm})$ we get that a minimum
 $E=1.3\times 10^{5}$~V/cm is required to induce the homogeneous
 state, a value well into the practical range.  To confirm our theory
 we propose the application of the $E$-field to bulk BFO along this
 specific direction.  The homogeneous state has as its optical
 signature the presence of only two magnon Raman modes
 \cite{desousa08b} (the signature of a canted antiferromagnet) instead
 of five or more cyclonic magnons \cite{cazayous08}.  We note that the
 usual largest side of BFO single crystals grown by the flux technique
 corresponds to the cubic $(010)$ plane. Thicker samples have to be
 grown and cut in order to select the appropriate direction.

 Some experiments seem to indicate the presence of homogeneous spin
 order in thin film samples of BFO \cite{bai05,bea07}.  Using a
 phenomenological theory, Bai {\it et al.} \cite{bai05} showed that
 the strain in films can destroy the cycloid state.  Our theory
 establishes a microscopic mechanism for destroying the cycloid in
 films that is unrelated to strain. In our model, the heterostructure
 inversion asymmetry leads to an internal $E$-field. A sufficiently
 large value of this field will induce homogeneous magnetic order.

 The ability to switch from cycloidal to homogeneous spin order
 \emph{without poling $\bm{P}$} is a pathway for $E$-field control of
 magnetism that avoids charge displacement and energy dissipation
 associated to the relaxation of $\bm{P}$ into another direction
 \cite{ashraf12}.  In BFO, the weak magnetization $\bm{M}\propto
 \bm{\hat{Z}}\times \bm{L}$ is tied to $\bm{L}$. Thus our mechanism
 allows the electrical switching of $\bm{M}$ from a sinusoidal state
 with zero spatial average to a homogeneous state with non-zero
 $\langle \bm{M}\rangle$.  This effect converts an $E$-field pulse
 into a magnetic pulse.  By combining BFO with another magnetic
 material (as done in \cite{heron11}), one can envision the writing of
 data in a magnetic memory element using an $E$-field pulse in an
 insulator instead of the usual current pulse in a metal.

 {\it Conclusions.---}We presented a microscopic theory of $E$-field
 induced magnetic anisotropy, and showed how it gives rise to an
 additional linear magnetoelectric effect (LME) in insulators.  The
 origin of this special kind of LME is based on the combination of two
 factors: The presence of a non-magnetic ion with large spin-orbit
 coupling, and a significant amount of inversion asymmetry (induced
 e.g. by ferroelectricity). For BFO, the presence of this additional
 LME implies that its magnetic cycloid can be converted into a
 homogeneous state under the application of a practical external
 $E$-field without the need for poling $\bm{P}$; and that the
 additional ferromagnetic degree of freedom $\bm{M}$ will be fully
 controllable by $\bm{E}$ and will not average out to zero over large
 length scales. Thus, it shows that $E$-field control of magnetism at
 room temperature can happen even without poling the ferroelectric
 polarization $\bm{P}$ into another direction, and can be done with
 much less energy dissipation than what has been demonstrated so far.

 Our research was supported by the NSERC Discovery program. The
 authors would like to thank D. Colson, I. Souza, and I. \v{Z}uti\'{c}
 for helpful discussions.


\begin{thebibliography}{99}
\bibitem{noteaxes} We used the definition
  $\bm{\hat{X}}=(-2\bm{\hat{x}}+\bm{\hat{y}}+\bm{\hat{z}})/\sqrt{6}$
  and $\bm{\hat{Y}}=(-\bm{\hat{y}}+\bm{\hat{z}})/\sqrt{2}$ as new
  rhombohedral axes perpendicular to
  $\bm{\hat{P}}=\bm{\hat{Z}}=(\bm{\hat{x}}+\bm{\hat{y}}+\bm{\hat{z}})/\sqrt{3}$.
\bibitem{noteq} The $E$-field necessary to flip the cycloid $\bm{q}$
  out of one of its three degenerate states ($\bm{q}\parallel
  [1,-1,0]$ and cyclic permutations) is $E_{\perp}\approx 0.6
  J(qR_{\rm{Bi}})^{4}/\xi\sim 10^{2}$~V/cm.
\end{thebibliography}
\end{document}